\documentclass[preprint,floatfix] {revtex4} 
 
\usepackage{graphicx}
\usepackage{subfigure}
\begin{document}

\title{Calculation of the spiked harmonic oscillators through a generalized
pseudospectral method}
\author{Amlan K. Roy}
\affiliation{Department of Chemistry, University of New Brunswick, Fredericton, NB, 
E3B 6E2, Canada}
\email{akroy@unb.ca}
\begin{abstract}
The generalized pseudospectral method is employed for the accurate calculation of 
eigenvalues, densities and expectation values for the  spiked harmonic 
oscillators. This allows \emph{nonuniform}
and \emph{optimal} spatial discretization of the corresponding single-particle
radial Schr\"odinger equation satisfying the Dirichlet boundary conditions leading 
to the standard diagonalization of the symmetric matrices. The present results 
for a large range of potential parameters are in excellent agreement with those from 
the other accurate methods available in the literature. The ground 
and excited states (both low as well as high angular momentum states) are obtained 
with equal ease and accuracy. Some new states including the higher excited states
are reported here for the first time. This offers a simple, accurate and efficient 
method for the treatment of these and a wide variety of other singular potentials of 
physical and chemical interest in quantum mechanics.
\end{abstract}

\maketitle

\section{Introduction}
A class of interaction potentials in quantum mechanics characterized by the 
Hamiltonian,
\begin{equation}
\mathrm{H}=\mathrm{p}^2+\mathrm{r}^2+\lambda |\mathrm{r}|^{-\alpha} 
\equiv \mathrm{H_0}+\lambda |\mathrm{r}|^{-\alpha}, \ \ 
\mathrm{r} \in [0, \infty]
\end{equation}
where $p=-i\ \partial /\partial \mathrm{r}$, 
have found widespread applications in many areas of atomic, molecular, nuclear physics
and are often referred as the spiked harmonic oscillators (SHO). Here $\mathrm{H_0}$ 
formally denotes
the simple harmonic oscillator Hamiltonian; the coupling parameter $\lambda$
determines the strength of the perturbative potential and the positive constant 
$\alpha$ represents the type of singularity at the origin. The higher the value of 
$\lambda$, the higher the singularity. There has been an upsurge of interest [1-20]
to calculate the SHO eigenvalues over
the past three decades and it continues to grow. An interesting feature of 
this potential is that once the perturbation $\lambda |\mathrm{r}|^{-\alpha}$ 
is switched on, complete turn-off is impossible; vestigial effects of the interaction 
persists leading to the so called ``Klauder phenomenon" [1,2]. From a purely mathematical 
viewpoint, on the other hand, this poses considerable challenges to some of the 
well-established and widely used mathematical theories. For example, the commonly used 
Rayleigh-Schr\"odinger perturbation series diverges [3] according to the relation 
$\mathrm{n} \geq \frac{1}{\alpha-2}$, where $\mathrm{n}$ is the order of the 
perturbation term. Consequently, a singular 
perturbation theory was to be specially devised to treat these potentials. These 
potentials also exhibit the phenomenon of {\it supersingularity} [1] in the region
of $\alpha \geq 5/2,$ i.e., every matrix element of the potential is infinite. 
The numerical solutions of the pertinent Schr\"odinger equations are notoriously 
difficult as well; especially those involving the finite-difference (FD) schemes
and often require special care. 

Several analytical (both variational and perturbative) methodologies [3-20] are  
available for the exact and approximate calculation of these systems.  
For example, the modified (nonpower) perturbation series [3] to finite order  
for the ground-state eigenenergies valid for small values of $\lambda$ and arbitrary
values of $\alpha$, a large coupling perturbative expansion [5] for the approximate 
estimates of the same for large positive values of $\lambda$, the weak coupling expansion
expressions of the nonsingular ($\alpha < 5/2$) SHO through
the resummation technique for $\alpha = 1/2, 1, 3/2\ $ [6] and for $\alpha = 2$ [10], 
the exact and approximate (variational) solutions [8] for some particular values of the 
parameters in the interaction potential, a modified WKB treatment [11], etc. Besides, 
the upper and lower bounds of ground and excited states [12-16] of the SHO as well as
the generalized SHO, the analytical 
pseudoperturbation shifted-$\ell$ expansion technique (PSLET) [19,20] have also been 
developed. The extensions to N-dimensions are reported lately [14,20]. On the numerical
side, the FD methods [21,22] through Richardson extrapolation, integration
of the Schr\"odinger equation [23] using a Lanczos grid method for the cases of
$\alpha = 4,6$ for small values of the ($\lambda \leq 0.01$), the analytic continuation
method [24] for ground and excited states have been reported. 

Despite all these attempts, a general prescription which can accurately and reliably
calculate the bound states of these potentials in a uniform and simple way for a general 
set of potential parameters with the promise of furnishing ground and excited states 
with equal ease, would be highly desirable. This is because
physically meaningful and good accuracy results are obtainable only by some of these
methods. Additionally some of these methods can give 
satisfactory results for a certain type of parameters while perform rather poorly in 
other cases. Much attention has been paid to the ground states; excited states are
reported less frequently and definitively, presumably because of the greater challenges  
compared to the ground states. Moreover, much work has been devoted to the eigenvalues;
only few results are available for the {\it eigenfunctions}[15]. Also some 
of these methodologies are often fraught with rather tedious and cumbersome 
mathematical complexities. This work proposes a simple methodology to study these
systems by using the GPS scheme which has shown considerable promise for a variety
of atomic and molecular processes including both static and dynamic situations in recent 
years (see, for example [25-29] and the references therein). This 
formalism helps alleviate some of the well-known discomfitures of the 
FD schemes widely used and discussed in the literature [5,21,22,30], 
e.g., the necessity of significantly larger spatial grid points to deal with 
the singularity at the origin. The GPS method essentially works in a {\it nonuniform}
and {\it optimal} spatial grid; thus a much smaller number of points suffices to
achieve good accuracy. However its applicability has been so far restricted to the 
cases of Coulomb singularities; no attempts have been made to deal with the other 
singularities characterizing many physical systems. The objective of this Letter is 
two-fold: (a) to extend the regions of applicability and judge the performance of it
on the SHOs, (b) to calculate accurately the bound-state spectra of these systems.
Comparison with the literature data has been made wherever possible. 

The organization of the article is as follows. Section II presents an overview of the 
basic formalism. Section III makes a discussion of the results while a few 
concluding remarks are made in section IV. 
 
\section{The GPS formalism for the SHO}\label{sec:method}
In this section, we present the GPS formalism used to solve the radial 
eigenvalue problem with the SHO potentials. A detailed account of the GPS method 
can be found in the references [25-29].

The time-independent radial Schr\"odinger equation to be solved can be written in the 
usual way (atomic units employed unless otherwise mentioned),
\begin{equation}
\left[-\frac{1}{2} \ \frac{\mathrm{d^2}}{\mathrm{dr^2}} + \frac{\ell (\ell+1)} 
{2\mathrm{r^2}} + \mathrm{v(r)} \right]
\psi_{n,\ell}(\mathrm{r})=E_{n,\ell}\ \psi_{n,\ell}(\mathrm{r})
\end{equation}
where v(r) is the SHO potential given by, 
\begin{equation}
\mathrm{v(r)}=[\mathrm{r}^2+\lambda/\mathrm{r}^{\alpha}]/2.
\end{equation}
The 1/2 factor is introduced here only for easy comparison 
with the literature and $\ell$ signifies the usual angular momentum quantum number. The 
GPS formalism facilitates the use of a denser mesh at small r regions and a relatively 
coarser mesh at the large r regions while preserving the similar accuracy at both the 
regions.

The key step in this formalism is to approximate a function $f(x)$ defined in the 
interval $x \in [-1,1]$ by an N-th order polynomial $f_N(x)$ {\it exactly} at the
discrete collocation points $x_j$ as in the following,  
\begin{equation}
f(x) \cong f_N(x) = \sum_{j=0}^{N} f(x_j)\ g_j(x),
\end{equation}
\begin{equation}
f_N(x_j) = f(x_j).
\end{equation}
Within the Legendre pseudospectral method that the current work uses, $x_0=-1$, 
$x_N=1$, and $x_j (j=1,\ldots,N-1)$ can be determined from the roots of the first 
derivative of the Legendre polynomial $P_N(x)$ with respect to $x$, i.e., 
\begin{equation}
P'_N(x_j) = 0.
\end{equation}
The $g_j(x)$s in Eq.~(4) are the cardinal functions expressed as,
\begin{equation}
g_j(x) = -\frac{1}{N(N+1)P_N(x_j)}\ \  \frac{(1-x^2)\ P'_N(x)}{x-x_j},
\end{equation}
satisfying the relation $g_j(x_{j'}) = \delta_{j'j}$.
At this stage, we use a transformation $r=r(x)$ to map the semi-infinite domain  
$r \in [0, \infty]$ onto the finite domain $x \in [-1,1]$. One can make use of the 
following algebraic nonlinear mapping,
\begin{equation}
r=r(x)=L\ \ \frac{1+x}{1-x+\alpha},
\end{equation}
where L and $\alpha=2L/r_{max}$ are the mapping parameters. Finally introduction of the 
following relation,
\begin{equation}
\psi(r(x))=\sqrt{r'(x)} f(x)
\end{equation}
in conjunction with a symmetrization procedure gives the following transformed Hamiltonian
\begin{equation}
\hat{H}(x)= -\frac{1}{2} \ \frac{1}{r'(x)}\ \frac{d^2}{dx^2} \ \frac{1}{r'(x)}
+ \mathrm{v}(r(x))+v_m(x),
\end{equation}
The advantage of this is that one ends up with a \emph {symmetric} matrix eigenvalue 
problem which can be solved readily and efficiently to give accurate eigenvalues 
and eigenfunctions by using standard routines. It may be noted that $v_m(x)=0$ for the 
above transformation leading to the following set of discretized coupled equations, 
\begin{widetext}
\begin{equation}
\sum_{j=0}^N \left[ -\frac{1}{2} D^{(2)}_{j'j} + \delta_{j'j} \ v(r(x_j))
+\delta_{j'j}\ v_m(r(x_j))\right] A_j = EA_{j'},\ \ \ \ j=1,\ldots,N-1,
\end{equation}
\end{widetext}
where
\begin{equation}
A_j  = \left[ r'(x_j)\right]^{1/2} \psi(r(x_j))\ \left[ P_N(x_j)\right]^{-1}.
\end{equation}
and the symmetrized second derivatives $D^{(2)}_{j'j}$ of the cardinal functions are 
given in [26]. Thorough checks are made on the variation of the energies with respect
to the mapping parameters for large ranges of the interaction parameters in the 
potential available in the literature. After a series
of such calculations, a choice has been made at the point where the results changed 
negligibly with any variation. In this way, a consistent and uniform set of parameters
($r_{max}=200,$ $\alpha=25$ and $N=300$) has been used. 

\section{Results and discussion}
\subsection{The charged harmonic oscillator, $\alpha=1$}
Before considering the general case of relatively stronger spikes, {\it viz.,} 
$\alpha \neq 1$, it is worthwhile to study the simpler special case of $\alpha=1$.
This does not exhibit supersingularity and the Hamiltonian takes the 
simplified confined Coulomb potential type form. It has been pointed out [31] that
this possesses an infinite set of {\em elementary} solutions. 
Table~\ref{tab:table1} displays such elementary solutions calculated by the present
method along with the exact analytical results. Note that $E=3/2$ is a trivial
solution corresponding to $\lambda=0$, i.e., the unperturbed Hamiltonian. The 
other $\lambda$s in this table are taken from the solutions of the polynomial 
equation [31]. It may be noted that all the calculated results in this table and 
throughout the article are {\em truncated} and therefore all the digits in the 
reported numbers should be considered as correct. It is seen that 
for all values of $\lambda$, our results match excellently up to a 12 digit accuracy 
with the exact values. 

\begingroup
\squeezetable
\begin{table}
\caption {\label{tab:table1}Some elementary solutions (in a.u.) of the SHO
with $\alpha=1$ for several values of $\lambda$ corresponding to the ground 
state.}
\begin{ruledtabular}
\begin{tabular}{ccc}
$\lambda$ & \multicolumn{2} {c} {Energy} \\ 
\cline{2-3} 
     & This work & Exact\footnote{Ref. [31]. These results have
been halved to take care of a 2 factor.}  \\  \hline
 0                          & 1.49999999999  &  1.5   \\
 2                          & 2.49999999999  &  2.5   \\
 $\sqrt{20} $               & 3.50000000000  &  3.5   \\
 $(30+6\sqrt{17})^{1/2}$    & 4.49999999999  &  4.5   \\
 $(70+6\sqrt{57})^{1/2}$    & 5.49999999999  &  5.5   \\
 14.450001026966            & 6.49999999999  &  6.5   \\
 18.503131410003            & 7.49999999999  &  7.5   \\
\end{tabular}
\end{ruledtabular}
\end{table}
\endgroup

\begingroup
\squeezetable
\begin{table}
\caption {\label{tab:table2}Calculated ground-state energies E (in a.u.) of the 
SHO with $\alpha=1$ for several values of $\lambda.$} 
\begin{ruledtabular}
\begin{tabular}{cccccl}
$\lambda$ & \multicolumn{2}{c}{Energy}& $\lambda$ & \multicolumn{2}{c}{Energy} \\ 
\cline{2-3} \cline{5-6}
         & This work & Literature\footnotemark[1]  &      & This work & 
Literature\footnotemark[1]     \\  \hline 
$-0.001$ & 1.49943577146 & 1.49943577146  & 0.001 & 1.50056415064 & 1.5005641506   \\
$-0.005$ & 1.49717807794 &                & 0.005 & 1.50281997477 &                \\
$-0.01$  & 1.49435420563 & 1.49435420565  & 0.01  & 1.50563800525 & 1.50563800525  \\ 
$-0.05$  & 1.47169265799 &                & 0.05  & 1.52811261097 &                \\
$-0.1$   & 1.44318757957 & 1.4431875796   & 0.1   & 1.55603345324 & 1.55603345325  \\ 
$-0.5$   & 1.20765362342 &                & 0.5   & 1.77283783394 &                \\
$-1$     & 0.892602739638 & 0.89260273965 & 1     & 2.02893850398 & 2.0289385040   \\ 
$-5$  & $-2.90807895034$ & $-2.90807895035$ & 5   & 3.69201586294 & 3.69201586295  \\
$-10$ & $-12.4404995301$ & $-12.44049953015$ & 10 & 5.28874176968 & 5.288741697    \\  
$-50$ & $-312.497600033$ &                 & 50   & 13.7025706824 &                \\
$-100$ & $-1249.99940000$ &                & 100  & 21.2314590573 &                \\ 
\end{tabular}
\end{ruledtabular}
\footnotetext[1]{Ref. [31]. The quoted results are halved to 
take care of a 2 factor.}
\end{table}
\endgroup

Next we report in Table~\ref{tab:table2} the ground-state energies for a large range
of (+)ve and (-)ve $\lambda$s (left and right sides of the table respectively) 
along with the available literature data. One can envisage three distinct regions 
in this case depending on the values of $\lambda$, {\em viz.,} (a) the Coulomb region, 
corresponding to large (-)ve $\lambda,$ (b) the strong-coupling region having large 
(+)ve $\lambda,$ and (c) the weak-coupling region having small (both (+)ve and 
(-)ve) $\lambda$. The perturbation expressions corresponding to regions (a) and 
(c) are obtained through an amalgamation of the hypervirial 
and Hellmann-Feynman theorem [31]. For some of the (-)ve and (+)ve $\lambda$s
ground states are examined by the Renormalization as well as the direct
numerical integration methods [31]. Also for $\lambda \leq -1$ and $\lambda \geq 1$, 
the Coulomb series and strong coupling series solutions are available [31].
Good agreement is observed for $\lambda=-10$ and $\lambda=10$ involving these methods;
for other $\lambda$s, they vary significantly from each other. Here, the numerical
results are quoted for comparison. No results were available for $\lambda = \pm 0.005, 
\pm 0.05, \pm 0.5, \pm 50, \pm 100)$. It is seen that the current results 
are in excellent agreement with theirs. At this point mention may be made of one of 
the uncomfortable features in some of the available methodologies, {\em viz.,} the
presence of the unphysical roots, e.g., in the Riccati-Pad\'e method for the small
$\lambda$s of these potentials [31]. However, no such solutions have been found 
in the present calculations. In some instances, very slight differences 
are observed in our results from the literature data. Furthermore, in 
table~\ref{tab:table3}, we present the calculated first three states 
corresponding to $\ell=0,1,2,3$ for these systems. The ground states are repeated
for the sake of completeness. Again a wide range of both positive and negative 
$\lambda$ values are chosen. No results are available for these states to our 
knowledge and we hope that they could be useful in future calculations. 

\begingroup
\squeezetable
\begin{table}
\caption {\label{tab:table3} Excited state energies (in a.u.) of the charged harmonic
oscillator for several positive and negative values of $\lambda$. First three states 
are presented corresponding to $\ell=0,1,2,3.$}
\begin{ruledtabular}
\begin{tabular}{cllll}
 $\lambda$  & $\ell=0$  &   $\ell=1$      &  $\ell=2$     &   $\ell=3$     \\ \hline 
 $-0.001$ & 1.49943577146   & 2.49962386468    & 3.49969909505    &  4.49974208263  \\
         & 3.49952982655   & 4.49966148110    & 5.49972058921    &  6.49975641177   \\
         & 5.49958155072   & 6.49968700564    & 7.49973670957    &  8.49976780986   \\
$-0.1$   & 1.44318757957   & 2.46229789284    & 3.46987169094    &  4.47418745667   \\
         & 3.45282982176   & 4.46609746678    & 5.47203352713    &  6.47562590482   \\
         & 5.45807015200   & 6.46866692538    & 7.47365242360    &  8.47676916627   \\
$-10  $  &$-12.4404995301$ &$-2.62119802134$  & 0.004574720607   & 1.67473695981    \\
         &$-2.41723883317$ & 0.551233459914   & 2.41797686878    & 3.89431134350    \\
         & 0.869699218970  & 3.039428778956   & 4.66996941607    & 6.05143674320    \\
$-100$   &$-1249.99940000$ &$-312.494000168$  &$-138.863694966$  &$-78.0530829912$  \\
         &$-312.491600236$ &$-138.852898182$  &$-78.0243305403$  &$-49.7760242855$  \\
         &$-138.847499606$ &$-78.0051588012$  &$-49.7312779249$  &$-34.2071382286$  \\
0.001    & 1.50056415063   & 2.50037611746    & 3.50030089728    & 4.50025791310    \\
         & 3.50047014252   & 4.50033850860    & 5.50027940556    & 6.50024358500    \\    
         & 5.50041843194   & 6.50031298742    & 7.50026328651    & 8.50023218755    \\
  0.1    & 1.55603345324   & 2.53752389333    & 3.53005208746    & 4.52577056764    \\
         & 3.54686142702   & 4.53380037959    & 5.52791522734    & 6.52434325783    \\
         & 5.54175768729   & 6.53126513099    & 7.52631014389    & 8.52320695048    \\
  10     & 5.28874176968   & 5.63241238009    & 6.19962012502    & 6.89697621559    \\
         & 7.07543947857   & 7.46149523127    & 8.06916090729    & 8.79599713087    \\
         & 8.89811648443   & 9.32030795656    & 9.96020685568    & 10.7102538225    \\
  100    & 21.2314590573   & 21.3064355531    & 21.4546955286    & 21.6730703270    \\
         & 22.9756496882   & 23.0536647120    & 23.2077138583    & 23.4341264159    \\
         & 24.7309414007   & 24.8119454465    & 24.9716696244    & 25.2059214990    \\
\end{tabular}
\end{ruledtabular}
\end{table}
\endgroup

\begingroup
\squeezetable
\begin{table} 
\caption {\label{tab:table4}Calculated ground-state energies E (in a.u.) of the SHO
with $\alpha=4$ and 6 for several values of $\lambda.$ The 
literature results are divided by a 2 factor.} 
\begin{ruledtabular}
\begin{tabular}{cllll}
$\lambda$ & \multicolumn{2}{c}{Energy ($\alpha=4$)} & 
\multicolumn{2}{c}{Energy ($\alpha=6$)} \\ 
\cline{2-3} \cline{4-5}
      & This work   & Literature   & This work    & Literature \\   \hline
0.001 & 1.53438158545 & 1.53438158545\footnotemark[1], 1.534385\footnotemark[2] & 
1.63992791296 & 1.63992791296\footnotemark[1]  \\
0.005 & 1.57417615416 & 1.574176155\footnotemark[3],1.574175\footnotemark[4],    & 
1.71144209213 & 1.71144208\footnotemark[3],1.71144\footnotemark[4],    \\
&     & 1.574195\footnotemark[5] &  & 1.71151\footnotemark[5] \\
0.01  & 1.60253374753 & 1.60253374753\footnotemark[1], 1.60254\footnotemark[2],  & 
1.75272613799 & 1.75272613799\footnotemark[1],1.752726195\footnotemark[3], \\ 
      &   & 1.602533745\footnotemark[3],1.602535\footnotemark[4],   &    & 
1.752725\footnotemark[4],1.75287\footnotemark[5],       \\
  &       & 1.602635\footnotemark[5], 1.602535\footnotemark[6]    &  
  &  1.7527265\footnotemark[6]       \\
0.05  & 1.71258069752 &                        & 1.88277010302 &          \\
0.1 & 1.78777599560 & 1.78777599560\footnotemark[1],
     1.787785\footnotemark[2] & 1.95783261264 &         \\ 
   &  & 1.787775\footnotemark[6]   &    &               \\
0.5   & 2.06529243634 &                        & 2.19395453013 &          \\
1     & 2.24708899168 & 2.24708899168\footnotemark[1],
     2.24709\footnotemark[2]$^,$\footnotemark[6]  & 2.32996998478 & 
2.32996998478\footnotemark[1],2.329970\footnotemark[6]  \\ 
5  & 2.89222177088 & 2.89222\footnotemark[6] 
   & 2.75657950709 & 2.7565795\footnotemark[6]               \\
10    & 3.30331125601 & 3.30331125601\footnotemark[1],
3.30331\footnotemark[2]$^,$\footnotemark[6]  & 3.00160451444 & 
3.00160451444\footnotemark[1],3.0016045\footnotemark[6], \\  
   &  & 3.3033112560\footnotemark[7]   &    &  3.00160451\footnotemark[7]  \\
50    & 4.73277787167 &                        & 3.76776072255 &         \\
100   & 5.63254021587 & 5.63254021587\footnotemark[1], 5.63254\footnotemark[2],  &
4.20667914031 & 4.2066791403\footnotemark[7] \\ 
    &   & 5.6325402\footnotemark[7]  &      &      \\
500   & 8.73793385806 &                        & 5.57607711626 &              \\
1000  & 10.6847312660 & 10.6847312660\footnotemark[1], 10.68473\footnotemark[2], & 
6.35930853290 &    \\
   &   &  10.684731265\footnotemark[7]   &    &    \\
\end{tabular}
\end{ruledtabular}
\footnotetext[1] {Ref. [24].}
\footnotetext[2] {Ref. [31].}
\footnotetext[3] {Ref. [23].}
\footnotetext[4] {Ref. [21].}
\footnotetext[5] {Ref. [1].}
\footnotetext[6] {Ref. [8].}
\footnotetext[7] {Ref. [33].}
\end{table}
\endgroup

\subsection{$\alpha \neq 1$}
Now results are presented for $\alpha \neq 1$. Here we focus on the $\alpha$ values
4 and 6; however,  the present scheme has been thoroughly checked to reproduce the 
results of similar accuracy and reliability for other values of $\alpha$
available in the literature. In table~\ref{tab:table4}, ground state energies are 
tabulated for these two cases ($\alpha=4$ in the left and $\alpha=6$ in the right), 
for small and large $\lambda$s. Two new $\lambda$ values 
are introduced here (500 and 1000) in addition to those employed in 
table~\ref{tab:table2}. Both of these $\alpha$ values can lead to supersingularity and 
have been investigated by many workers.  The present results are seen to be in good 
agreement with the accurate analytic continuation results [24]. These results are 
available for $\lambda= 0.001, 0.01, 1, 10$ for both $\alpha=4,6$, while 
$\lambda=0.1, 100, 1000$ for $\alpha=4$ only. Various other results are also 
available for the smaller $\lambda$s (0.005, 0.01) [23,21,1,33] 
and our results show good agreement with these. Direct integration results 
[8] for $\lambda=0.01, 0.1, 1, 5, 10$ are also presented for comparison. It may be 
noted that the current results surpass in accuracy to all others except [24].

\begingroup
\squeezetable
\begin{table} 
\caption {\label{tab:table5}Calculated $\ell \ne 0$ state energies (in a.u.) of the 
SHO with $\alpha=4$ (top) and 6 (bottom) for several values of $\lambda$s. The
literature results have been divided by a 2 factor.} 
\begin{ruledtabular}
\begin{tabular}{cccccc}
$\ell$ & $\lambda=0.001$ & $\lambda=0.01$ & $\lambda=0.1$ & $\lambda=1$ &
$\lambda=10$ \\ \hline
 3 & 4.50005713956 & 4.50057109970 & 4.50568201308 & 4.55432930375 & 4.91961566042  \\
   & 4.50005713956\footnotemark[1] & 4.50057109970\footnotemark[1] & 
     4.50568201309\footnotemark[1] & 4.55432930376\footnotemark[1] &                \\
 4 & 5.50003174537 & 5.50031739444 & 5.50316804961 & 5.53112085969 & 5.77200022575  \\
   & 5.50003174537\footnotemark[1] & 5.50031739444\footnotemark[1] & 
     5.50316804961\footnotemark[1] & 5.53112085969\footnotemark[1] &                \\
 5 & 6.50002020182 & 6.50020200030 & 6.50201821626 & 6.52000759152 & 6.68566506197  \\ 
   & 6.50002020182\footnotemark[1] & 6.50020200030\footnotemark[1] & 
     6.50201821626\footnotemark[1] & 6.52000759153\footnotemark[1] &                \\ 
10 & 11.5000050125 & 11.5000501247 & 11.5005011983 & 11.5050070693 & 11.5495902896  \\
20 & 21.5000012507 & 21.5000125077 & 21.5001250765 & 21.5012506189 & 21.5124915772  \\
30 & 31.5000005556 & 31.5000055570 & 31.5000555707 & 31.5005556887 & 31.5055549875  \\
40 & 41.5000003125 & 41.5000031254 & 41.5000312547 & 41.5003125438 & 41.5031249904  \\
50 & 51.5000001999 & 51.5000020001 & 51.5000200019 & 51.5002000183 & 51.5020000374  \\
   &               &               &               & 51.5002000183\footnotemark[1] & \\
\hline 
5  & 6.50000577192 & 6.50005771227 & 6.50057643602 & 6.50570148892 & 6.55258902874  \\ 
10 & 11.5000005897 & 11.5000058970 & 11.5000589687 & 11.5005894939 & 11.5058757159  \\
20 & 21.5000000675 & 21.5000006760 & 21.5000067609 & 21.5000676086 & 21.5006759799  \\
30 & 31.5000000194 & 31.5000001949 & 31.5000019498 & 31.5000194985 & 31.5001949796  \\
40 & 46.5000000080 & 41.5000000811 & 41.5000008117 & 41.5000081181 & 41.5000811806  \\
50 & 51.5000000040 & 51.5000000411 & 51.5000004123 & 51.5000041240 & 51.5000412410  \\
\end{tabular}
\end{ruledtabular}
\footnotetext[1] {Ref. [16].}
\end{table}
\endgroup

Now table~\ref{tab:table5} gives the results for low- and high-$\ell$ states for a wide
range of $\lambda$s (0.001, 0.01, 0.1,1,10) for both $\alpha=4,6$. Upper and 
lower bounds as well as the numerical eigenenergies for $\ell=3,4,5$ have been studied
recently by [16] for $\alpha=4$ for first four $\lambda$s. Our results match almost 
completely with theirs except very slight discrepancies in three instances at the 
last digit (our results are lower by $10^{-11}$). Also, the eigenvalues of 
$\ell=5,10,20 \cdots ,50$ for all the mentioned values of $\alpha$ 
and $\lambda$ are given as a test of this method for the very high excited states. 
The present result is in complete agreement with the lone available result of 
$\ell=50$ (for $\alpha=4, \lambda=1$). Next table~\ref{tab:table6} gives results 
for the first 10 eigenvalues of the SHO with the parameters $\alpha=6, \lambda=10$. 
We have considered $\ell=0,1,2,3,4$ and no results could be found for these states.  
 
\begingroup
\squeezetable
\begin{table}
\caption{\label{tab:table6}The first 10 eigenvalues (in a.u.) for $\ell=0,1,2,3,4,$
of the SHO. The parameters are: $\alpha=6$ and $\lambda=10.$}
\begin{ruledtabular}
\begin{tabular}{ccccc}
$\ell=0$      &  $\ell=1$      & $\ell=2$       & $\ell=3$       & $\ell=4$  \\  \hline
 3.00160451444 & 3.32389487858 & 3.91806927392 & 4.70345973112 & 5.60034573904  \\
 5.38666914834 & 5.64859182812 & 6.14941053052 & 6.84505383654 & 7.67818492569  \\
 7.66493489996 & 7.89510635043 & 8.34183646535 & 8.97697716408 & 9.75876666401  \\
 9.88940298242 & 10.0989808012 & 10.5089815278 & 11.0996720737 & 11.8395020670  \\
 12.0805100442 & 12.2752417097 & 12.6580500545 & 13.2142245371 & 13.9191144855  \\
14.2485304976 & 14.4318512146 & 14.7933959369 & 15.3217108425 & 15.9969892830  \\
16.3994450470 & 16.5736092280 & 16.9178854110 & 17.4230621225 & 18.0728546082  \\
18.5370785045 & 18.7036642068 & 19.0335218023 & 19.5190555671 & 20.1466196426  \\
20.6640433547 & 20.8242086533 & 21.1417690049 & 21.6103351695 & 22.2182898578  \\
22.7822131369 & 22.9368391875 & 23.2437342203 & 22.6974360875 & 24.2879213038  \\
\end{tabular}
\end{ruledtabular}
\end{table}
\endgroup

\begingroup
\squeezetable
\begin{table} 
\caption {\label{tab:table7}Calculated expectation values (in a.u.) for the SHO
for some selected values of $\alpha\ $ and $\lambda\ $. The first three states 
corresponding to $\ell=0,1,2\ $ are presented.} 
\begin{ruledtabular}
\begin{tabular} {ccccc}
$\alpha$ &$\lambda$ &$\ell$ &$\langle r^{-1}\rangle$ & $\langle r \rangle $ \\ \hline 
 1   & 10      &   0     & 0.579335567 & 1.88860444    \\
     &         &         & 0.572186022 & 2.20351385    \\     
     &         &         & 0.562374825 & 2.49562513    \\
 4   & 10      &   0     & 0.546623313 & 1.93946889    \\
     &         &         & 0.483512472 & 2.38064959    \\     
     &         &         & 0.443751364 & 2.74044976    \\
 6   & 10      &   0     & 0.558986259 & 1.89176957    \\
     &         &         & 0.477245223 & 2.38688068    \\     
     &         &         & 0.431013450 & 2.77224304    \\
\end{tabular}
\end{ruledtabular}
\end{table}
\endgroup

As a test on the quality of the eigenfunctions, In table~\ref{tab:table7}, we present
some of the calculated expectation values $\langle r^{-1} \rangle$ and 
$\langle r \rangle $ for $\alpha=1,4$ and 6. The parameter 
$\lambda$ is kept fixed at 10 in all these cases and the first three states are 
reported for $\ell=0,1,2$. No results could be found for any of these 
values in the literature. Finally, figure 1 depicts the radial probability 
distribution functions for the first three states of $\ell=0,1,2$ along 
with the potential ($\alpha=6, \lambda=10$). As expected they show the 
requisite number of nodes in these plots. 

\begin{figure}
\begin{minipage}[b]{0.40\textwidth}
\centering
\includegraphics[scale=0.30]{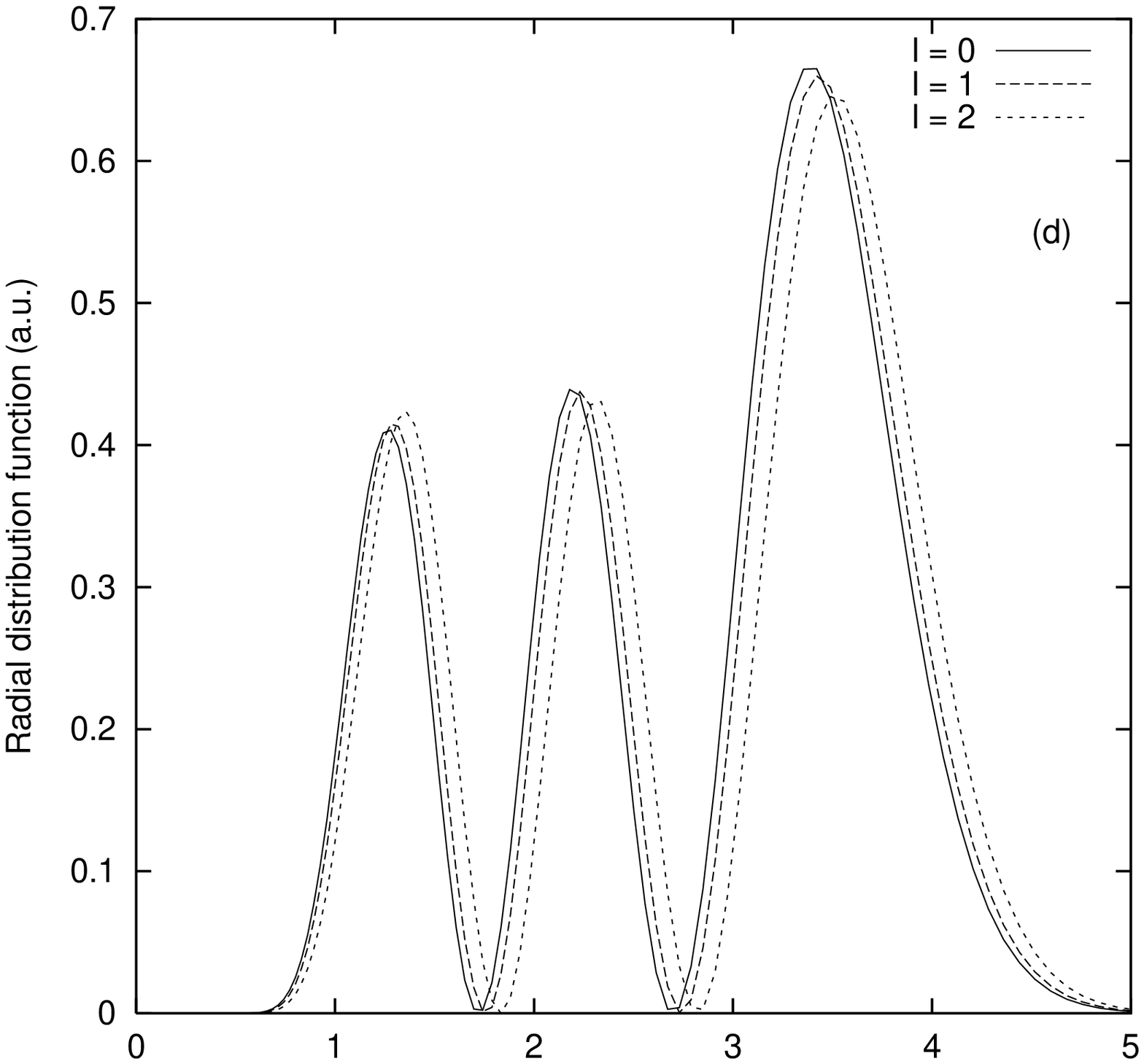}
\end{minipage}%
\\
\begin{minipage}[b]{0.40\textwidth}
\centering
\includegraphics[scale=0.30]{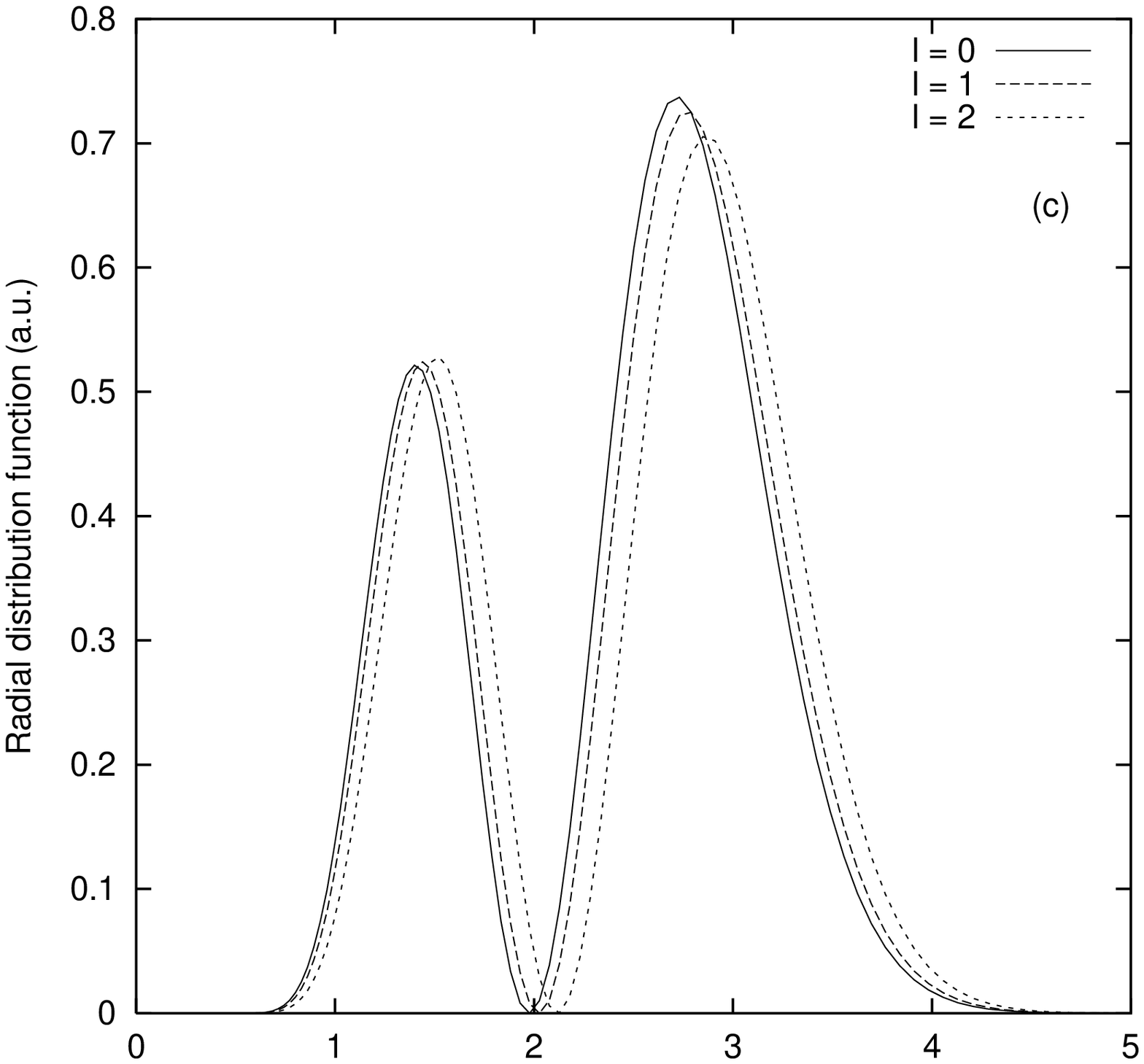}
\end{minipage}%
\\
\begin{minipage}[b]{0.40\textwidth}
\centering
\includegraphics[scale=0.30]{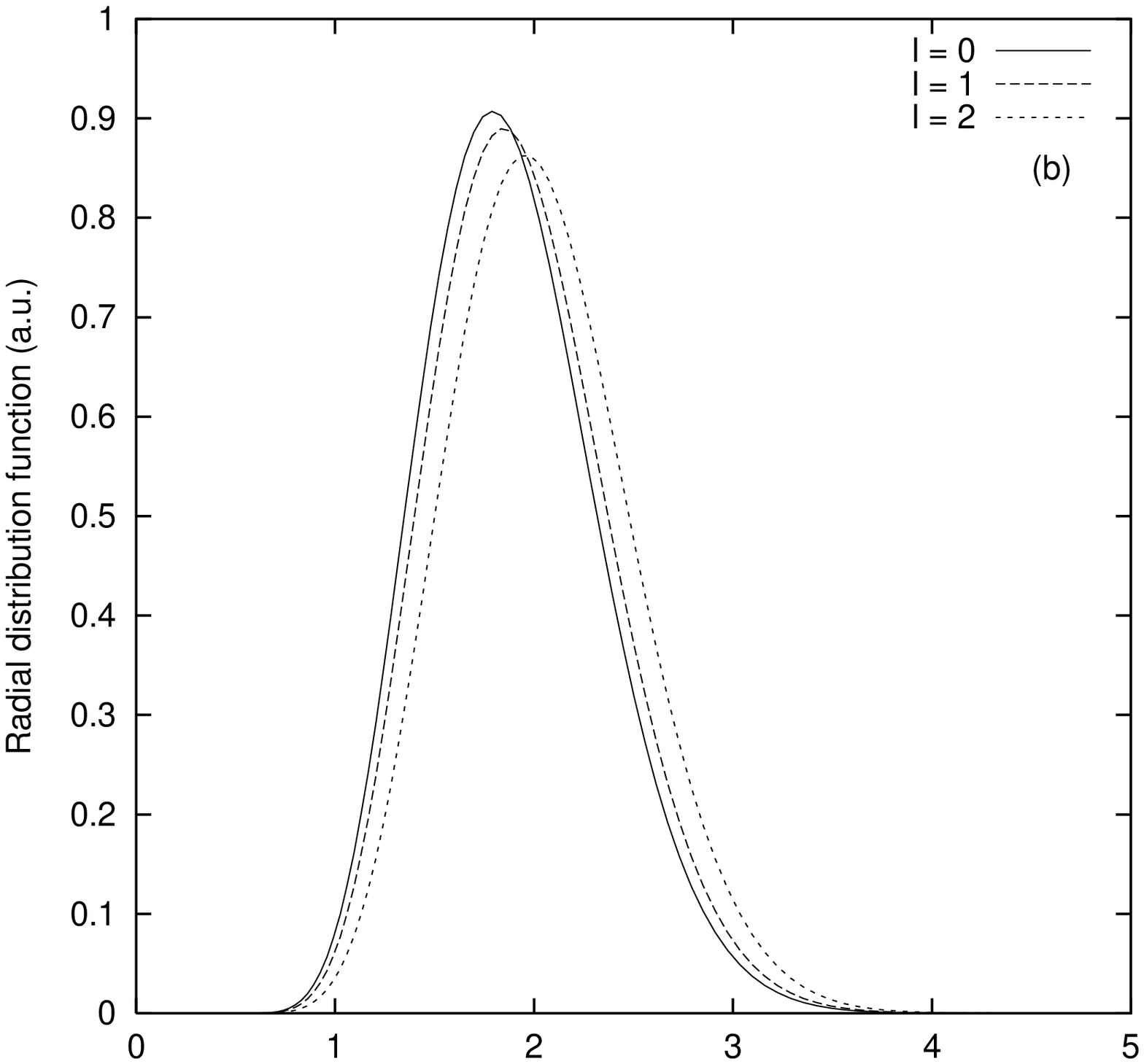}
\end{minipage}%
\\
\begin{minipage}[b]{0.40\textwidth}
\centering
\includegraphics[scale=0.30]{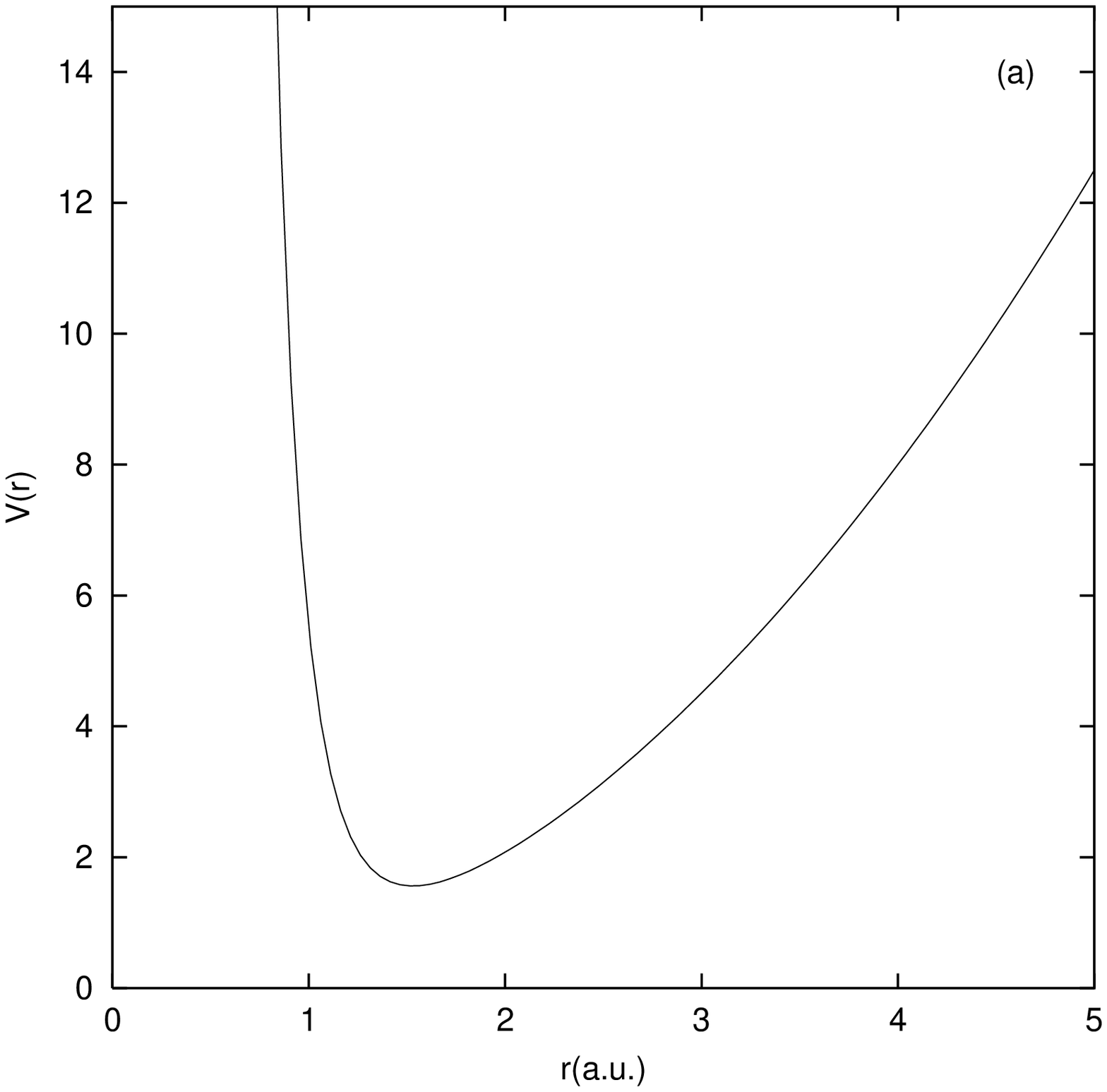}
\end{minipage}%

\caption{The radial probability distribution function, $|rR_{n\ell}|^2$ (in a. u.)  for
the first three states corresponding to $\ell=0,1,2$ of the spiked oscillator potential 
having the parameters, $\alpha=6,\ \lambda=10.\ $ (a) The potential, (b) the ground 
state, (c) the first excited state and (d) the second excited state.}\label{fig:fig1}
\end{figure}

\section{Conclusion}
The GPS formalism is shown to deliver {\em accurate} and {\em reliable} results for 
the eigenvalues, expectation values and the radial densities of the SHOs. The 
simplicity and viability of the method is demonstrated by calculating the low and 
high excited states of these potentials for weak and strong values of the interaction 
parameter in the potential. Excellent agreement with the literature data is observed 
in all cases. Some states are reported here for the first time. Finally the approach 
may be as well equally successful and useful for other singularities (e. g., the 
Hulth\'en, Yukawa, Hellman potentials etc.) in quantum mechanics. Work in this direction is
under progress. 

\begin{acknowledgments}
I gratefully acknowledge the hospitality provided by the Department of Chemistry, 
University of New Brunswick, Fredericton, Canada. 
\end{acknowledgments}

\end{document}